\documentstyle[12pt]{article}

\newcommand{\lbl}[1]{\label{eq:#1}}

\newcommand{\vs}[1]{\rule[- #1 mm]{0mm}{#1 mm}}
\newcommand{\eq}{\vs{2}\begin{equation}}
\newcommand{\en}{\\[2mm]\end{equation}}
\newcommand{\bea}{\begin{eqnarray}}
\newcommand{\ena}{\end{eqnarray}}
\newcommand{\NP}[1]{Nucl.\ Phys.\ {\bf #1}}

\newcommand{\PL}[1]{Phys.\ Lett.\ {\bf #1}}

\newcommand{\AN}[1]{Ann. Phys. {\bf #1}}

\newcommand{\PRev}[1]{Phys.\ Rev.\ {\bf #1}}
\newcommand{\PRL}[1]{Phys.\ Rev.\ Lett.\ {\bf #1}}

\begin{document}
\setcounter{page}{0}
\renewcommand{\thefootnote}{\fnsymbol{footnote}}

\rightline{IPNO/TH 93-17}

\indent

\indent

\begin{center}
{\Large {\bf A possible experimental determination of $m_s /{\hat m}$ }}

{\Large {\bf from $K_{\mu 4}$ decays}}

\vskip 3 true cm

{\large M. Knecht, H. Sazdjian, J. Stern }

\indent

{\sl Division de Physique Th\'eorique
\footnote{Unit\'e de Recherche des Universit\'es Paris XI et Paris VI
associ\'ee au CNRS.}, Institut de Physique Nucliaire\\
F-91406 Orsay Cedex, France} \\

\indent

{\large N.H. Fuchs}

\indent

{\sl Department of Physics, Purdue University, West Lafayette IN 47907}

\end{center}

\vskip 4 true cm

\centerline{\large {\bf Abstract}}

\indent

$K\pi$ scattering and $K_{\mu 4}$ decays are studied at leading order of
improved chiral perturbation theory. It is shown that high precision
$K_{\mu 4}$ experiments at, e.g., DA$\Phi$NE should
allow for a direct measurement of the quark mass ratio $m_s /{\hat m}$.

\indent

\indent

\noindent IPNO/TH 93-17 \\
April 1993
\newpage


\renewcommand{\thefootnote}{\arabic{footnote}}
\setcounter{footnote}{0}


\indent

\indent

The light flavour symmetry breaking sector of QCD involves various parameters
whose precise determination is of a fundamental importance. In particular, the
products of running quark masses $m_u$, $m_d$, $m_s$ with the quark-antiquark
condensate of the massless theory,
\eq
B_0 \equiv  - {F_0^{-2}} <{\bar u}u>
    = - {F_0^{-2}} <{\bar d}d>
    = - {F_0^{-2}} <{\bar s}s>\ ,
\en
are well defined renormalization group invariant quantities which are in
principle measurable and which are not determined within the standard model.
($F_0$ denotes the chiral limit of the pion decay constant $F_{\pi}$ = 93.1
MeV.) While quark masses can be chosen freely within QCD, the magnitude of the
scale dependent condensate $B_0$ is an intrinsic property of the theory,
reflecting the mechanism of spontaneous symmetry breaking. Since the latter is
not yet clearly understood in QCD, the order of magnitude of $B_0$ is hard to
estimate {\it a priori}:
$B_0$ could be as large \cite{GOR} \cite{QM} as the scale ${\Lambda}_H$ of
formation of massive bound states, ${\Lambda}_H \sim$ 1 GeV, or it could be
as small as the fundamental order parameter $F_0 \sim$ 90 MeV.

For sufficiently small quark masses, the expansions of Goldstone boson masses
are dominated by the linear terms
\eq
M_{\pi^+}^2 = (m_u + m_d )B_0 + ...\ ,\
 M_{K^+}^2 = (m_u + m_s )B_0 + ...\ ,\
 M_{K^0}^2 = (m_d + m_s )B_0 +...\ .
\en
How small the quark masses should actually be in order to ensure this dominance
is controlled by the size of $B_0$: For the pseudoscalar meson ${\bar a}b$
($a$, $b$ = $u$, $d$, $s$; $a\ne b$) such a dominance requires
\eq
m_a + m_b \ll B_0 / A_0 \ ,
\en
where $A_0$ is a dimensionless parameter of order one characteristic of
contributions to the expansion (2) coming from terms which are quadratic in
quark masses. ($A_0$ has been defined in Ref.
\cite{FSS1} in terms of a two-point QCD correlator.) For $B_0$ of the order of
the bound state scale $\Lambda_H \sim$ 1 GeV, the condition (3) is likely to
be satisfied for actual values of quark masses. In this case, the standard
\cite{GL} chiral perturbation theory ($\chi$PT) - which counts each insertion
of quark mass as {\it two} powers of pion momentum - should describe the low
energy data well already within a few lowest orders. If, on the other
hand, $B_0$ turned out to be comparable to the fundamental order parameter
$ F_0\sim$ 90 MeV, the condition (3) would certainly be violated already for
$m_a$ or $m_b$ equal to the strange quark mass, but also for non strange
quark masses in the range 20 - 30 MeV, where they are still small as compared
to $\Lambda_H$. The first term in the expansion (1)
would then be considerably lower than the pseudoscalar masses $M_P^2$, and
consequently, the standard expansion of the symmetry breaking part of the QCD
effective lagrangian should be rearranged in order to improve its convergence.
An improved $\chi$PT has been proposed in references \cite{FSS2}, \cite{FSS3}:
 it is an expansion in pion momentum $p/{\Lambda_H}$, in quark masses
${\cal M}_q /{\Lambda_H}$ and in powers of $B_0 /{\Lambda_H}$, with
${\cal M}_q$ and $B_0$ counting as a {\it single} power of $p$. This modified
counting rule leads to a consistent redefinition of individual chiral orders.
The leading $O(p^2 )$ order now consists of 5 independent terms: in a standard
notation,
\bea
{\tilde{\cal L}}_{2} &=& {F_0^2\over 4}\big\{
                   \langle D^{\mu}U^+ D_{\mu}U\rangle +
        2B_0 \langle {\cal M}_q U + {\cal M}_q U^{+}\rangle\nonumber\\
        &+& A_0 \langle {({\cal M}_q U)}^2 + {({\cal M}_q U^{+})}^2 \rangle +
        Z_0^S{\langle {\cal M}_q U + {\cal M}_q U^{+}\rangle}^2\\
   &+& Z_0^P{\langle{\cal M}_q U - {\cal M}_q U^{+}\rangle}^2 \big\}\nonumber
\ .\lbl{imp}
\ena
The terms quadratic in the quark mass matrix ${\cal M}_q$ that are usually
 relegated to the $O(p^4 )$ order \cite{GL} can now give contributions
comparable to the $B_0$-term, reflecting the violation of condition (3).
(Notice that $Z_0^S$ and $Z_0^P$ violate the Zweig rule in the $0^+$ and $0^-$
channels, respectively. $Z_0^P$ will play no role in the present work.)

The improved $\chi$PT generalizes the standard expansion since at each order
the former contains additional terms, that the latter relegates to higher
orders. Consequently, it is less predictive, but
constitutes a more appropriate theoretical framework for an unbiased
experimental determination of symmetry breaking parameters such as the ratios
of quark masses, $m_q B_0$ and other non-perturbative characteristics of the
massless QCD vacuum. It is convenient to use the improved $O(p^2 )$ expression
for $M_{\pi}^2$ and $M_K^2 = {1\over 2}(M_{K^+}^2 + M_{K^0}^2 )$,
\bea
&&M^2_{\pi} = 2{\hat m}{\mu}_0 + 4{\hat m}^2 A_0\ ,\nonumber\\
&&{}\lbl{mass}\\
&&M^2_K = ({\hat m} + m_s){\mu_0} + {({\hat m} + m_s)}^2 A_0\ ,\nonumber
\ena
where ${\hat m} = {1\over 2}(m_u + m_d )$,
 ${\mu}_0 = B_0 + 2(m_s + 2{\hat m}) Z_0^S$, and to express the low energy
constants of ${\tilde{\cal L}}_2$ in terms of
\eq
r = {{m_s}\over{{\hat m}}}\ \quad ,\ \ {\zeta} = {{Z_0^S}\over{A_0}}\ .
\en
(In a similar way, the constant $Z_0^P$ can be expressed in terms of the
${\eta}$ mass.) The masses $M_{\pi}^2$, $M_K^2$, the quark mass ratio $r$ and
the Zweig rule violating constant $\zeta$ are independent parameters, except
for the restriction \cite{FSS1}
\eq
r_1 \equiv 2\,{{M_K}\over{M_{\pi}}} - 1 \le r \le r_2 \equiv
2\,{{M_K^2}\over{M_{\pi}^2}} - 1\ ,
\en
arising form the requirement of vacuum stability. The leading order of the
standard $\chi$PT \cite{GL} is reproduced for the particular choice $r = r_2
\sim$ 25.9, $Z_0^S =$ 0, implying $A_0 =$ 0. The other extreme, viz. $r = r_1
\sim$ 6.3, $\zeta =$ 0, corresponds to the order parameter $B_0 =$ 0.
The value of the quark mass ratio $r$ should ultimately be determined from
experiment, which may comfirm or invalidate the {\it a priori} estimate
$r\sim r_2$. Actually, a recent analysis of the deviations from the
Goldberger-Treiman relation suggests that $r$ might be less than 25 by a
factor of 2 or 3 \cite{FSS1}.

There are not many physical processes directly accessible to experiment that
exhibit a strong dependence on the quark mass ratio $r$ already at the order
$O(p^2 )$. One of them is the $\pi - \pi$ scattering. At the tree level, the
corresponding amplitude can be parametrized as
\eq
A(s\vert t, u) = {{\beta}_{\pi\pi}\over{F_{\pi}^2}}(s - {4\over3}M_{\pi}^2 )
                           + {{\alpha}_{\pi\pi}\over{3 F_{\pi}^2}}M_{\pi}^2\ ,
\en
where, at leading $O(p^2 )$ order \cite{FSS2}, \cite{FSS3},
\eq
{\alpha}^{lead}_{\pi\pi} = 1 + 6\,{{r_2 - r}\over{r^2 - 1}}\, (1 +
2\zeta)\quad\quad ,\quad
{\beta}^{lead}_{\pi\pi} = 1 \ .
\en
As $r$ decreases from $r_2$ to $r_1$, ${\alpha}^{lead}_{\pi\pi}$ increases
from the canonical value ${\alpha}^{lead}_{\pi\pi} =$ 1 \cite{W} to
${\alpha}^{lead}_{\pi\pi} =$ 4. A method of determining $r$ from the
forthcoming precise low-energy $\pi - \pi$ scattering data has been discussed
in Ref. \cite{FSS3}.

The main purpose of this letter is to point out that there exists another
independent case of similar interest: the decay
\eq
{K^+} \to {\pi}^{+} {\pi}^{-} {\mu}^{+}{{\nu}_{\mu}}\ \ ,
\en
which, although less abundant than the standard $K_{e4}$ decay, can be easily
accessible at future high statistics Kaon factories, e.g. at DA$\Phi$NE
\cite{DHB}. To the leading $O(p^2 )$ order, the axial-vector part of the
$K_{l4}$ matrix element receives two contributions, shown in Fig.1. While the
direct interaction vertex of Fig.1a is independent of $r$, the $K-$pole
contribution of Fig.1b exhibits an $r$ dependence through the virtual $\pi -
K$ scattering amplitude. The latter gives rise to a contribution proportional
to the lepton mass, and hence invisible in $K_{e4}$ decays. In this paper, the
question of whether the $r$ dependence can be observed in the $K_{\mu 4}$
decays
is answered positively within the leading order. The loop corrections to
this result will be presented elsewhere.

\indent

\indent

Ignoring isospin breaking effects (from now on $m_u = m_d = {\hat
m}$), the amplitude for the $\pi - K$ scattering process
\eq
{\pi}^a + K^i \to {\pi}^b + K^j \ ,
\en
$a, b = 1, 2, 3,\  i, j = \pm 1/2$, is described in terms of two invariant
amplitudes ${A^{\pm}}(s, t, u)$,
\eq
A^{{\pi}^a + K^i \to {\pi}^b + K^j}(s, t, u)\  =\
     {\delta}^{ab}{\delta}^{ij} A^{+}(s, t, u) - i {\epsilon}^{abc}
          {\big({\tau}^c\big)}^{ij} A^{-}(s, t, u) \ ,
\en
with
\eq
A^{\pm}(s, t, u) = {\pm} A^{\pm}(u, t, s) \ .
\en
They are related to the isospin amplitudes $A^{I} (s, t, u),\ I = {1\over 2},
{3\over 2}$, by
\bea
&&A^{3/2}  = A^{+} + A^{-} \nonumber\\
&&{}\\
&&A^{1/2}  = A^{+} - 2A^{-} \nonumber \  .
\ena
Upon neglecting $O(p^4)$ terms, the tree level amplitudes are described by
three constants: ${\alpha}_{\pi K}$, ${\beta}_{\pi K}$, analogous to the
constants ${\alpha_{\pi\pi}}$ and ${\beta_{\pi\pi}}$ occurring in the low
energy parametrization of the ${\pi}-{\pi}$ tree level amplitude in Eq.(8),
and ${\gamma}_{\pi K}$,
\eq
A^{+}(s, t, u) = {{\beta}_{\pi K}\over{4 F^{2}_{\pi}}}
         (t - {2\over 3} M_{\pi}^2 - {2\over 3} M_{K}^2 )
       + {1\over {6 F^{2}_{\pi}}}\big\{ ( M_K - M_{\pi})^2 +
2M_{\pi}M_K{\alpha}_{\pi K}\big\}
\en
\eq
A^{-}(s, t, u) = {{\gamma}_{\pi K}\over{4 F_{\pi}^2}} (s - u)\ .
\en
At leading order, these constants read
\eq
{\alpha}_{\pi K}^{lead} - 1 = {{r + 1}\over{r_1 + 1}}\,
(\alpha_{\pi\pi}^{lead} - 1)\quad,\ \
{\beta}_{\pi K}^{lead} = 1 = {\gamma}_{\pi K}^{lead}\ \ .
\en
At this stage, the $r$ dependence enters the $K-{\pi}$ amplitude through the
constant ${\alpha}_{\pi K}$ only (cf. the similar situation in the
${\pi}-{\pi}$ case). When $r$ differs from $r_2$ this leads to an enhancement
of
the $K-{\pi}$ amplitude: as it is defined, $\alpha_{\pi K}^{lead}$, like
$\alpha_{\pi\pi}^{lead}$, varies from 1 (the standard case \cite{W},
\cite{BKM}) to 4, for $r =  r_1$. At the same order, the scattering lengths
$a_l^I$ and slope parameters $b_l^I$ ($l$ = 0, 1, $I$ = ${1\over{2}}$,
${3\over{2}}$) read \footnote{For the definition of the threshold parameters
$a_l^I$ and $b_l^I$ we follow the conventions of Ref. \cite{GL}}
\bea
&&a^{1/2}_0  = {{M_{\pi}M_K}\over{32\pi F_0^2}}\  {{5 + \alpha_{\pi
K}^{lead}}\over{3}}\ ,\nonumber\\
&&a^{1/2}_1  = {1\over{64\pi F_0^2}}\ ,\nonumber\\
&&a^{3/2}_0  = -{{M_{\pi}M_K}\over{32\pi F_0^2}}\  {{4 - \alpha_{\pi
K}^{lead}}\over{3}}\ ,\nonumber\\
&&a^{3/2}_1  = 0\ , \\
&&b^{1/2}_0  = {-1\over{32\pi F_0^2}}\,\bigg[\, {3\over2} - {{(M_{\pi} +
M_{K})^2}\over {{M_{\pi}}{M_{K}}}}\, \bigg]\ ,\nonumber\\
&&b^{3/2}_0  = {-1\over{32\pi{F_0^2}}}\  {{(M_{\pi} +
M_{K})^2}\over {{M_{\pi}}{M_{K}}}}\ .\nonumber
\ena
One notices that the combination $2a_0^{3/2} + a_0^{1/2}$, which
vanishes in the standard case, is the most sensitive one to departures of $r$
from $r_2$. The combination $a_0^{3/2} - a_0^{1/2}$, which may,
in principle, be determined through an accurate measurement of the lifetime of
$K - \pi$ atoms \cite{N}, does not depend on $r$ at leading order. (The
lifetime of $\pi - \pi$ atoms similarly gives access to the combination
$a_0^2 - a_0^0$ \cite{N} of $\pi - \pi$ scattering lengths, which still depends
on $r$ at leading order \cite{FSS3}.) On the other hand, model independent
informations on the $I$ = ${1\over{2}}$ phase shifts may be extracted from high
precision data on $D_{l4}$ decays (e.g. $D^+ \to K^+ {\pi}^- e^+ {\nu}_e$).
Quite generally, and as already noticed
in the case of $\pi - \pi$ scattering \cite{FSS3}, for $r \sim$ 10, the
improved leading order modifies the scattering lengths in the same direction
and by roughly the same amount as the standard loop corrections \cite{GL},
\cite{BKM}.

\indent

\indent

Next, we turn to the $\, K_{l4}\, $ decays, $\, l=e,\, \mu$; we shall
concentrate on the process
\eq
{K^+}(k) \to {\pi}^{+}(p_{+}) {\pi}^{-}(p_{-}) l^{+}({p_l}){{\nu}_l}(p_{\nu})
\ .
\en
The axial current matrix element is described by three form factors, $F$, $G$,
$R$,
\bea
&&<{\pi}^{+}(p_{+}){\pi}^{-}(p_{-})\vert A^{4-i5}_{\mu} \vert K^{+}(k)>
=\quad\nonumber\\
&&\\
&&\qquad   = {-i\over{M_K}}\ \bigg[\, {(p_{+} + p_{-})}_{\mu}\, {F} +
                         {(p_{+} - p_{-})}_{\mu}\, {G} +
                         {(k - p_{+} -p_{-})}_{\mu} {R}\, \bigg]\ ,\nonumber
\ena
while the vector current matrix element requires only one form factor, $H$,
\eq
<{\pi}^{+}(p_{+}){\pi}^{-}(p_{-})\vert V^{4-i5}_{\mu} \vert K^{+}(k)> =
   - {{H}\over{M^3_K}}\, {\epsilon}_{\mu\nu\rho\sigma}\, k^{\nu}
    \, {(p_{+} + p_{-})}^{\rho}{(p_{+} - p_{-})}^{\sigma}\ .
\en
These form factors are functions of the invariants
\bea
&& s_{\pi} = {(p_{+} + p_{-})}^2\ ,\nonumber\\
&& s_l \, = {(k - p_{+} -p_{-})}^2\ ,\\
&& \Delta \, = -2k\cdot (p_{+} - p_{-})\ .\nonumber
\ena
Contributions to ${H}$ start only at order $O(p^4)$ in the effective
lagrangian with the Wess-Zumino term which gives
\eq
H = -\ {{{\sqrt 2} M_K^3}\over{8{\pi}^2 F_0^2}}\ .
\en
At leading order, the form factors ${F}$ and ${G}$ are also constant and read
\eq
F\ =\ G\ =\ {M_K\over{{\sqrt 2}{F_0}}}\ .
\en
The form factor $R$ is the sum
\eq
R = R_{direct} + R_{K-pole}\ .
\en
$R_{direct}$ arises from diagrams where the axial current $A_{\mu}^{4-i5}$
couples directly to three pseudoscalar mesons (Fig. 1a), while $R_{K-pole}$ is
obtained from diagrams where $A^{4-i5}_{\mu}$ couples to a single internal
pseudoscalar line (Fig. 1b). At leading order, one obtains
\eq
R_{direct} = {M_K\over{{\sqrt 2}F_0}}\cdot{2\over 3}\ ,
\en
and
\eq
R_{K-pole} = {M_K\over{{\sqrt 2}F_0}}\cdot {1\over{s_l  - M_K^2}}\,\bigg\{
                  {1\over 2}s_{\pi} + {1\over 2}\Delta -
                  {1\over 6}(s_l - M_K^2 ) + {2\over 3} M_{\pi}M_K\,
                    ({\alpha}_{\pi K}^{lead} - 1)\,\bigg\}\ .
\en
At leading order, the dependence on $r$ appears only in $R_{K-pole}$, through
the $K-{\pi}$ scattering parameter ${\alpha}_{\pi K}^{lead}$.
In the differential decay rate, the contributions of $R$ appear always with a
multiplicative factor $m_l^2$ (a review of the kinematics of $K_{l 4}$
processes and explicit formulae for differential decay rates may be found in
Refs. \cite{BEG}, \cite{B}).
Hence, $K_{e 4}$ decays will be quite insensitive to the value of $r$. On the
other hand, $K_{\mu 4}$ processes offer the possibility for a direct
experimental determination of $m_s/{\hat m}$. In Fig. 2, we have plotted the
differential decay rate $d\Gamma / ds_l$ for different values of $r$ using the
$O(p^2 )$ expressions for the form factors $F$, $G$, $R$, and formula (23) for
$H$. As $r$ varies between $r_1$ and $r_2$, one sees an overall effect of 20 -
25 $\%$, and which is not sensitive to the value of the Zweig rule violating
parameter $\zeta$ taken in the range 0 - 0.2. (One could, in principle, obtain
both the values of $r$ and $\zeta$ from separate measurements of
$\alpha_{\pi\pi}$ and of $\alpha_{\pi K}$.) A statistical sample of 30.000
events, which might be obtained at DA$\Phi$NE, should be sufficient for
an experimental determination of $r$. At order $O(p^2 )$, one computes the
total decay rate for the process (19) to be $\Gamma =$ 156 s$^{-1}$ for $r =
r_2$ and $\Gamma =$ 112 s$^{-1}$ for $r = r_1$.
The loop corrections are expected to modify the above results in a
nonnegligible way. There is however no reason to believe that they would
destroy the sensitivity with respect to $r$ exhibited at leading order. The
experience from the $\pi - \pi$ analysis shows that loops rather tend to
amplify the tree level effects. The results of the loop calculations will be
presented elsewhere.

\indent

\newpage
\noindent{\Large{\bf Figure Captions}}

\indent

\indent

\noindent {\bf Figure 1}. The matrix element of the axial current
$A_{\mu}^{4-i5}$ (wavy line) between an incoming $K$ (solid line) and two
outgoing $\pi$'s (broken lines), showing: a) the direct contribution and, b)
the $K$-pole contribution to the form factor $R$.

\indent

\indent

\noindent {\bf Figure 2}. The differential $K_{\mu 4}$ decay rate
${d\Gamma}/{ds_l}$ (in units of $M_{\pi}^{-1}$) as a function of $s_l$ (in
units of $M_{\pi}^2$) plotted for different values of $r$ and for $\zeta =$
0.1. Starting from the bottom curve, correponding to $r$ = $r_1\sim$ 6.3, the
subsequent curves correspond to $r$ = 10, $r$ = 15 and $r$ = $r_2\sim$ 25.9,
respectively.
\end{document}